\documentclass[aoas,MSNbibl,nameyear,seceqn,dvips]{arximspdf}
\usepackage{dcolumn}
\usepackage{graphicx}

\doi{10.1214/13-AOAS686} 
\volume{8}
\issue{1}
\pubyear{2014}
\firstpage{286}
\lastpage{308}

\makeatletter
\newcolumntype{d}[1]{D{.}{.}{#1}}
\newcommand{\sign}{\operatorname{sign}}
\newproclaim{pro}{Procedure}
\newtheorem{teo}{Theorem}[section]
\makeatother

\begin{document}
\begin{frontmatter}

\title{Applying multiple testing procedures to detect change in East African vegetation}
\runtitle{Detecting change in East African vegetation}

\begin{aug}
\author[A]{\fnms{Nicolle} \snm{Clements}\corref{}\thanksref{t1,m1}\ead[label=e1]{nclement@sju.edu}},
\author[B]{\fnms{Sanat K.} \snm{Sarkar}\thanksref{t2,m2}\ead[label=e2]{sanat@temple.edu}},
\author[B]{\fnms{Zhigen}~\snm{Zhao}\thanksref{t3,m2}\ead[label=e3]{zhaozhg@temple.edu}}
\and
\author[C]{\fnms{Dong-Yun} \snm{Kim}\thanksref{m3,m4}\ead[label=e4]{kimd10@nhlbi.nih.gov}}
\runauthor{Clements, Sarkar, Zhao and Kim}
\affiliation{Saint Joseph's University\thanksmark{m1},
Temple University\thanksmark{m2},
National Institute of Health\thanksmark{m3}\break and
Virginia Tech\thanksmark{m4}}
\address[A]{N. Clements\\
Saint Joseph's University\\
Mandeville 349\\
5600 City Ave\\
Philadelphia, Pennsylvania 19131\\
USA\\
\printead{e1}}
\address[B]{S. K. Sarkar\\
Z. Zhao\\
Temple University\\
Speakman 331\\
1810 North 13th Street\\
Philadelphia, Pennsylvania 19122-6083\\
USA\\
\printead{e2}\\
\phantom{E-mail:\ }\printead*{e3}}
\address[C]{D.-Y. Kim\\
National Institute of Health\\
Office of Biostatistics Research\\
6701 Rockledge Drive\\
Bethesda, Maryland 20892\\
USA\\
\printead{e4}} 
\end{aug}
\thankstext{t1}{Supported by NSF Grant DMS-10-06344.}
\thankstext{t2}{Supported by NSF Grants DMS-10-06344 and DMS-12-08735.}
\thankstext{t3}{Supported by NSF Grant DMS-12-08735.}

\received{\smonth{1} \syear{2013}}
\revised{\smonth{9} \syear{2013}}

%
\begin{abstract}
The study of vegetation fluctuations gives valuable information toward
effective land use and development. We consider this problem for the
East African region based on the Normalized Difference Vegetation Index
(NDVI) series from satellite remote sensing data collected between 1982
and 2006 over 8-kilometer grid points. We detect areas with significant
increasing or decreasing monotonic vegetation changes using a multiple
testing procedure controlling the mixed directional false discovery
rate (mdFDR). Specifically, we use a three-stage directional
Benjamini--Hochberg (BH) procedure with proven mdFDR control under
independence and a suitable adaptive version of it. The performance of
these procedures is studied through simulations before applying them to
the vegetation data. Our analysis shows increasing vegetation in the
Northern hemisphere as well as coastal Tanzania and generally
decreasing Southern hemisphere vegetation trends, which are consistent
with historical evidence.
\end{abstract}

%

\end{frontmatter}

\section{Introduction}\label{sec1}

The need to understand the Earth's ecology and land cover is becoming
increasingly important as the impacts of climate change start to affect
animal and plant life, which ultimately affect human life. Knowledge of
current vegetation trends and the ability to make accurate predictions
is essential to minimize times of food scarcity in underdeveloped
countries. Vegetation trends are also closely related to sustainability
issues, such as management of conservation areas and wildlife habitats,
precipitation and drought monitoring, improving land usage for
livestock, and finding optimum agriculture seeding and harvest dates
for crops.

The United Nations has given attention recently to precipitation and
vegetation monitoring in East Africa, where a severe drought hit the
entire region in mid-2011. The drought has caused a food crisis across
Somalia, Ethiopia and Kenya, threatening the livelihood of over 10
million people. In many areas, the precipitation rate during the
``long'' rainy season from April to June 2011 was less than 30\% of the
average of 1995--2010. The lack of rain led to vegetation decline, crop
failure and widespread loss of livestock, as high as 40\%--60\% in some
areas [\citet{OCHA2011}].

Droughts are commonly thought to occur from prolonged periods with less
than average precipitation, which is then followed by a decline in
vegetation growth. However, droughts can also arise independently from
precipitation changes when soil conditions and erosion are triggered by
poorly planned agricultural endeavors. Overfarming, excessive
irrigation and deforestation can all adversely impact the ability of
the land to capture and hold water. Thus, current efforts are geared
toward monitoring agriculture and vegetation changes in hopes of
minimizing the effects of low precipitation rates in future years.

Assessment of changes in a region's vegetation structure is
challenging, especially in topographically diverse areas, like East
Africa. Forecasting future vegetation and agricultural planning become
particularly difficult when unknown trends are occurring. However, the
regions with vegetation changes are often the areas of most interest in
land use management. For example, if a previously underdeveloped region
is experiencing increasing trends in vegetation growth, meaning the
land is able to sustain plant growth, local farmers could utilize this
area to grow crops or raise livestock in future years. On the other
hand, if a region is experiencing decreases in vegetation growth, this
could be an indicator of overfarming, putting crops and livestock at
risk of drought.

Data collection on vegetation and land cover are typically done through
satellite remote sensing. The remote sensing imagery is used to convert
the observed elements (i.e., the image color, texture, tone and
pattern) into numeric quantities at each pixel in the image. The image
pixels correspond to a square grid of land, the size of which depends
on the satellite's resolution. One such numeric value is the normalized
difference vegetation index (NDVI). The NDVI has been shown to be
highly correlated with vegetation parameters such as green-leaf biomass
and green-leaf area, and hence is of considerable value for vegetation
monitoring [\citet{Curran1980},
\citet{JacksonSlaterPinter1983}]. The NDVI standard scale ranges
from $-$1 to 1, indicating how much live green vegetation is contained
in the targeted pixel. An NDVI value close to 1 indicates more abundant
vegetation. Low values of NDVI (say, 0.1 and below) correspond to
scarce vegetation consisting mostly of rock, sand and dirt, for
example. A range of moderate values (0.2 to 0.3) indicates short
vegetations such as shrub or grassland; larger NDVI values can be found
in rainforests (0.6 to 0.8). Often, negative NDVI values are
consolidated to be zero since negative values indicate nonvegetation
and are of little use for vegetation monitoring.

Statistical and computational methods are needed to analyze remotely
sensed data, like NDVI values, to determine trends in land condition
and to predict areas at risk from degradation. Methodologies that
detect land cover changes need to be sensitive as well as accurate,
since it can be costly and risky to relocate human populations,
agriculture or livestock to new regions of detected change. In such
spatio-temporal data, existing change detection methodologies include
geographically weighted regression [\citet{Foody2003}], principal
component analysis [\citet{HayesSader2001}] and smoothing
polynomial regression [\citet{ChenJonssonTamura2004}]. However,
these methods are unable to provide an upper bound on false detections.
Since there is large risk associated with falsely declaring an area to
have significant vegetation changes, land use managers seek new methods
that have a meaningful control over such errors.

In this article, we revisit the problem of detecting vegetation changes
in East Africa based on the NDVI data and propose applying multiple
testing methodologies. Such methodologies are very useful for detecting
changes of statistical significance with a control over an overall
measure of false detections and are being currently used in many other
modern scientific investigations. We should point out that
\citet{VrielingdeBeursBrown2008} did investigate this problem in a
hypothesis testing framework, but, unlike ours, did not attempt to
address the inherent multiplicity issue by controlling an overall false
detection rate while making their final conclusions. Our proposed
multiple testing methods have been developed by fine-tuning some
existing ones in order to adequately capture the specific data
structure and answer questions in the present context. In particular,
there is a local dependency among nearby hypotheses (e.g., neighboring
NDVI pixels) that should be taken into account and one should be able
to identify the increasing/decreasing direction of a vegetation trend
for an NDVI pixel once a significant change is detected. Our methods
aim to incorporate such local dependency and control an error rate, the
mixed directional false discovery rate (mdFDR), which is an overall
measure of nondirectional as well as directional false detections.

We organize the paper as follows. In Section~\ref{sec2} we describe the
East African NDVI data, its source and the associated multiple testing
problem. In Section~\ref{sec3} we present our proposed mdFDR
controlling procedures after providing some background information and
notation related to multiple testing, and assessing spatial
correlation. We consider dividing the East African region into
subregions to adequately capture local dependencies among the NDVI
values. The semivariogram plot [\citet{Cressie2011}], which is
used to investigate the presence of spatial autocorrelation, helps
determine the size of each subregion. We propose two procedures,
Procedures \ref{pro1}~and~\ref{pro2}, to control the mdFDR under such
group or block \mbox{dependence} structure. Procedure~\ref{pro1} is
referred to as a three-stage directional Benjamini--Hochberg (BH)
procedure whose mdFDR control is theoretically shown (in an
\hyperref[app]{Appendix}) assuming independence between but not within
blocks (or subregions) and numerically examined under various
dependence scenarios through simulations. Procedure~\ref{pro2} is an
adaptive version of Procedure~\ref{pro1} designed to improve
Procedure~\ref{pro1} through estimating the proportion of true null
hypotheses within each subregion, and its mdFDR control is studied only
through simulations. The findings of these simulations are reported in
Section~\ref{sec4}. In Section~\ref{sec5} we illustrate the
applications of these proposed methods to the NDVI data collected in
East Africa from 1982--2006. Discussions and concluding remarks are in
Section~\ref{sec6}.

\section{Data description and the statistical problem}\label{sec2}

East Africa spans a wide variety of climate types and precipitation
regimes which are reflected in its vegetation cover. To capture this,
satellite imagery was collected over a sub-Saharan region of East
Africa that includes five countries in their entirety (Kenya, Uganda,
Tanzania, Burundi and Rwanda) and portions of seven countries (Somalia,
Ethiopia, South Sudan, Democratic Republic of Congo, Malawi, Mozambique
and Zimbabwe). This roughly ``rectangular'' region extends from
27.8$^\circ$E to 42.0$^\circ$E longitude and 15$^\circ$S to
6.2$^\circ$N latitude. Also included in the region are several East
African Great Lakes such as Lake Victoria, Lake Malawi and Lake
Tanganyika.

The remotely sensed images were recorded twice a month from 1982--2006
and then converted to NDVI values. Hence, the spatio-temporal data set
consists of approximately 50,000 sites (pixels), each with 600 time
series observations (24 observations per year over 25 years). The
satellite's resolution corresponds to each pixel spanning an 8~km
${\times}8$~km grid of land. This Global Inventory Modeling and Mapping
Studies (GIMMS) data set is derived from imagery obtained from the
Advanced Very High Resolution Radiometer (AVHRR) instrument onboard the
National Oceanic and Atmospheric Administration (NOAA) satellite series
7, 9, 11, 14, 16 and 17. The NDVI values have been corrected for
calibration, view geometry, volcanic aerosols, cloud coverage and other
effects not related to vegetation change
[\citet{TuckerPinzonBrownSlaybackPakMahoneyVermoteSaleous2005}].

%
\begin{figure}[b]
\includegraphics{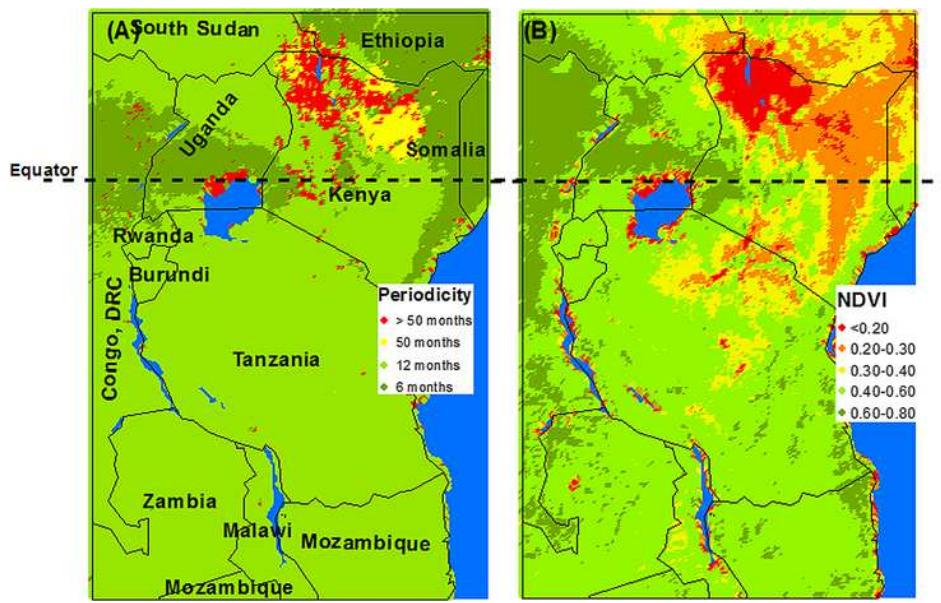}
\caption{\textup{(a)}~Periodicity categories of the NDVI time series for
the time period from 1982--2006. \textup{(b)}~Average NDVI for the time
period of 1982--2006 in East Africa.}\label{fig1}
\end{figure}

\begin{figure}[b]
\includegraphics{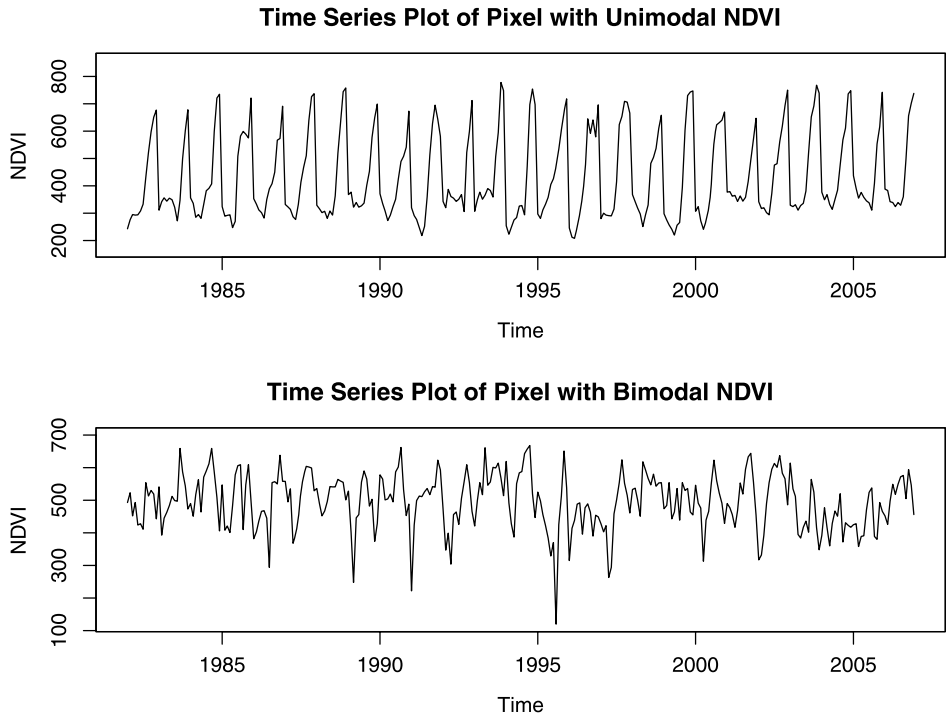}
\caption{The time series plot for two selected pixels: a unimodal
periodic pattern (top) and a bimodal periodic pattern (bottom). The
top is a plot of a pixel in South Sudan (31.0$^\circ$E, 5.6$^\circ$N), while the bottom series represents a pixel in Rwanda
(30.1$^\circ$E, 1.9$^\circ$S).}\label{fig2}
\end{figure}

The cyclic/seasonal behavior of the NDVI time series at each pixel in
the region is color-coded in Figure~\ref{fig1}(a). For example, areas with dark
green have a six-month periodicity while areas with light green have a
twelve-month periodicity. Periodogram analysis indicates very strong
peaks at six and twelve months, respectively, in these regions. The
periodicities are reflected in bimodal and unimodal shapes in annual
NDVI series, which in turn correspond to two and one rainy season each
year. Figure~\ref{fig1}(b) displays the average NDVI values for each grid point
(site) over the region. Blue areas indicate regions containing only
water (Indian Ocean, Lake Victoria, etc.), and thus no vegetation index
was recorded. The light and dark green areas have more green vegetation
on average compared to the drier areas, represented with yellow, orange
and red. In this figure, one can see how this East African region spans
the NDVI scale. Desert regions (with low NDVI) are within a few hundred
kilometers of wetlands and rain forests (with extremely high NDVI),
illustrating the large variability of climate types and precipitation
regimes in this region. Figure~\ref{fig2} shows the time series plots for two
selected pixels. The top series is a pixel selected from Southern Kenya
and has a unimodal periodic pattern. The bottom series of Figure~\ref{fig2} is a
pixel selected from the Democratic Republic of the Congo and has a
bimodal periodic pattern.

We consolidated all the negative NDVI values to zero, as commonly done
in vegetation monitoring, and then rescaled the remaining values by
1000. This is because negative values indicate nonvegetation areas, so
they are of little use for our purpose. Prior to the analysis, we
examined the data for quality assurance and eliminated a small number
of pixels that were found to have several consecutive years with
identical data values, which may be due to data entry errors or machine
malfunction.

This data set was first examined in
\citet{VrielingdeBeursBrown2008} where the interest was in
studying several phenology indicators, including start of the season,
length of season, time of maximum NDVI, maximum NDVI and cumulative
NDVI over the season. After extracting these indicators for every year,
trend tests were conducted to detect regions of significant changes in
phenology indicators. The percentage of pixels with the trend test
$p$-value less than $\alpha= 0.10$ for each phenology indicator was
reported separately for positive and negative slopes. The reported
results indicate that much of the region has ``significant'' vegetation
change. For example, the cumulative NDVI indicator detected 44.2\% of
sites with $p$-values less than 0.10. However, this study fails to
address the important statistical issue of multiplicity when making
these claims about significant vegetation changes and their directions
simultaneously for all the regions based on hypothesis testing.

When testing a single null hypothesis against a two-sided alternative,
two types of error can occur when a directional decision is made
following rejection of the null hypothesis. These are Type~I error and
Type~III (or directional) errors. The Type~I error occurs when the null
hypothesis is falsely rejected, while the Type~III error occurs when
the null hypothesis is correctly rejected but a wrong directional
decision is made about the alternative. For instance, when declaring a
particular 8~km${}\times{}$8~km grid of land as ``significantly''
changing in terms of vegetation, a~Type~I error is made if the area is
not truly changing, and a Type~III error is made if the area is truly
changing but in the opposite direction of what is determined from the
data. When such decisions are made simultaneously based on testing
multiple hypotheses, as in \citet{VrielingdeBeursBrown2008}, one
should adjust for multiplicity and control an overall measure of Types
I~and~III errors. Without such multiplicity adjustment, more Types
I~and~III errors can occur than the desired $\alpha$ level. It is
particularly important to avoid these errors as much as possible in the
present application. Land use managers, government and local farmers
are looking to relocate East African populations of people, livestock
and crops to areas of promising vegetation changes and avoid regions
with decreasing changes. Since these migrations can be risky and
costly, a careful consideration of the multiplicity issue seems
essential when making declarations of significant vegetation changes.

In this paper, we revisit the work in
\citet{VrielingdeBeursBrown2008} to adequately address the
multiplicity issue. To test each 8~km${}\times{}$8~km grid of land for
vegetation change, we use the cumulative NDVI phenology indicator for
each season or, equivalently, use the average NDVI per season. Since
the East African region straddles the Equator, ``seasons'' are
classified by precipitation changes rather than temperature, and it is
quite probable that a particular site can have vegetation changes in
one season and not another. This region receives rain in two distinct
seasons, locally referred to as the ``long rains'' (April--June) and
the ``short rains'' (November--December). The long rains provide more
rainfall than the short rains, but generally the arrival of the short
rains is more predictable. In hopes of capturing any seasonal changes,
trend tests were conducted on the seasonal NDVI averages at each site
for the first dry season (January--March), long rain season
(April--June), second dry season (July--October) and short rain season
(November--December).

To test for significant trend in each of the four seasons, we apply the
monotonic trend test proposed by \citet{Brillinger1989} for a time
series consisting of a signal and stationary autocorrelated errors. We
use the seasonal averages for each year as the observed time series.
This test examines the null hypothesis that the series has a~signal,
that is, constant in time against the alternative hypothesis that the
signal is monotonically increasing or decreasing in time. The test
statistic is a standardized version of a linear combination of the time
series values, with coefficients given in
\citet{AbelsonTukey1963}. This statistic is approximately normal
with mean zero if and only if the null hypothesis is true. More
specifically, given the 25-year data $\bar{Y}_{\mathrm{FD}}(t)$,
$t=0,1, \ldots, 24$, on the first dry (FD) season NDVI average in a
particular site, Brillinger's trend test can be applied for that season
assuming the model.
%
%
\begin{equation}
\bar{Y}_{\mathrm{FD}}(t) = S_{\mathrm{FD}}(t) +
E_{\mathrm{FD}}(t)
\end{equation}
for $t=0,1, \ldots, 24$, where $ S_{\mathrm{FD}}(t)$ is a deterministic
signal, and $E_{\mathrm{FD}}(t)$ is a zero mean stationary noise
series, for $t =0,1, \ldots, 24$. The test statistic is the ratio of
the linear combination $\sum_{t=0}^{24} c(t) \bar{Y}_{\mathrm{FD}}(t)$,
with
\[
c(t) = \biggl\{t \biggl(1- \frac{t}{25} \biggr) \biggr\}^{1/2} -
\biggl\{(t+1) \biggl(1- \frac{t+1}{25} \biggr) \biggr\}^{1/2},
\]
$t=0,1, \ldots, 24$, and the estimate of the standard error of this
linear combination. The hypotheses of interest are the null
$H_{0,\mathrm{FD}}\dvtx \beta_{\mathrm{FD}} = 0$ and the two-sided
alternative $H_{1,\mathrm{FD}}\dvtx \beta_{\mathrm{FD}} \neq0$, where
$\beta_{\mathrm{FD}} = \sum_{t=0}^{24} c(t)S_{\mathrm{FD}}(t)$. This
test can be similarly applied for testing the vegetation trend for the
remaining three precipitation seasons. These tests were implemented by
adapting R code written by Dr. Vito Muggeo, found at
\texttt{%
\href{https://stat.ethz.ch/pipermail/r-help/2002-December/027669.html}{https://stat.ethz.ch/}
\href{https://stat.ethz.ch/pipermail/r-help/2002-December/027669.html}{pipermail/r-help/2002-December/027669.html}}.

Thus, for each site (8~km${}\times{}$8~km grid of land), we have four
$p$-values, each providing an evidence of vegetation change occurring
over the years in that particular season---the smaller the $p$-value,
the higher is the evidence of a significant vegetation change. Our goal
is to do the following for each site: (i)~combine the four seasonal
$p$-values to form a yearly $p$-value, (ii) decide based on this yearly
\mbox{$p$-}value if a significant vegetation change has occurred over
the years at that site, and (iii) if vegetation change is found
significant, detect the season(s) that contributes to this change as
well as the direction in which this change has taken place. We wish to
accomplish this goal simultaneously for all sites ($\approx$50,000) in
the East African region in a multiple testing framework designed to
ensure a control over a meaningful combined measure of statistical
Types I~and~III errors.

\section{The proposed multiple testing procedures}\label{sec3}
We propose two multiple testing procedures that would be suitable for
applications to the vegetation data. Before that, we need to provide
some background in multiple testing and a brief outline of our idea to
determine the size of each subregion capturing local spatial
dependencies using variograms.

\subsection{Background in multiple testing}\label{sec3.1}
When simultaneously testing several null hypotheses, procedures have
traditionally been developed to control the familywise error rate
(FWER), which is the probability of at least one Type~I error (i.e.,
rejecting true null hypothesis), at a desired level, say, $\alpha$.
However, this notion of error rate is often too conservative when the
number of hypotheses being tested becomes large, as in the present
application. Therefore, there has been a recent surge in statistical
research to define alternative, less stringent error rates and to
develop multiple testing methods that control them. The false discovery
rate (FDR), which is the expected proportion of Type~I errors among all
rejected null hypotheses, introduced by
\citet{BenjaminiHochberg1995}, is one of these alternative error
rates that has received much attention.

In \citet{BenjaminiHochberg1995} a method was proposed, known as
the BH method for short, for controlling the FDR. For testing $m$ null
hypotheses $H_i$, $i=1, \ldots, m$, using their respective $p$-values
$P_i$, $i=1, \ldots, m$, it operates as follows: consider the ordered
versions of the $p$-values, $P_{(1)} \le\cdots\le P_{(m)}$, find
$k=\max\{i\dvtx P_{(i)} < \frac{i \alpha}{m} \}$, and reject the null
hypotheses whose $p$-values are less than or equal to $P_{(k)}$,
provided the maximum exists; otherwise, accept all null hypotheses.
This procedure controls the FDR at level $\alpha$, under the assumption
of \mbox{independence} or positive dependence (in a certain sense) of
the $p$-values. More specifically, the FDR of the BH method equals
$\pi_0 \alpha$ when the $p$-values are independent, and is less than
$\pi_0 \alpha$ when the $p$-values are positively dependent
[\citet{BenjaminiYekutieli2001}, \citet{Sarkar2002}], where
$\pi_0$ is the (true) proportion of null hypotheses. The difference
between $\pi_0 \alpha$ and the FDR gets larger with increasing
(positive) dependence among the $p$-values.


Often, it becomes essential for researchers, as in the present
application, to determine the direction of significance, rather than
significance alone, when testing multiple null hypotheses against
two-sided alternatives. In other words, for each test, researchers have
to decide whether or not the null hypothesis should be rejected and, if
rejected, determine the direction of the alternative. Typically, this
direction is determined based on the test statistic falling in the
right- or left-hand side of the rejection region. Such decisions can
potentially lead to one of two types of errors for each test, resulting
in rejection of the null hypothesis---the Type~I error if the null
hypothesis is true or the directional error, also known as the Type~III
error, if the null hypothesis is not true but the direction of the
alternative is falsely declared.

To deal with both Types I~and~III errors in an FDR framework, the
notion of mixed directional FDR (mdFDR) has been introduced in
\citet{BenjaminiYekutieli2005}. It is defined as the expected
proportion of Types I~and~III errors among all rejected null
hypotheses. A method in \citet{BenjaminiYekutieli2005} was given
for independent tests that controls the mdFDR when testing multiple
simple hypotheses against two-sided alternatives. They proved that the
original BH method controlling the FDR at $\alpha$ can be augmented to
make directional decision upon rejecting a null hypothesis according to
the corresponding test statistic falling in the right- or left-hand
side of the rejection region without causing the mdFDR to exceed
$\alpha$. This so-called augmented method is referred to as the
directional BH procedure.

This directional BH procedure will be extended in this paper to a
situation, conforming more to the present application, where the
$p$-values are not all independent but can be grouped in such a way
that they are mostly dependent within but not between the groups.
Unlike in the case of multiple testing applications to genomics where
gene pathways provide a natural way of grouping the $p$-values, there
are no clearly defined so-called ``vegetation pathways'' in the present
application that we can consider for grouping the $p$-values.
Nevertheless, a statistically meaningful approach can be devised for
grouping $p$-values using variograms as outlined in the following section.

\subsection{Variogram and its use in forming subregions capturing local
spatial dependencies}\label{sec3.2}

The variogram is an important characteristic describing the degree of
spatial dependence of a spatial random field or stochastic process $ \{
Z(\mathbf{s})\dvtx  \mathbf{s} \in D \}$. The variogram is defined as
$2\gamma(\mathbf{s},\mathbf{h}) = \operatorname{Var} [ Z(\mathbf{s})-
Z(\mathbf{s}+\mathbf{h)}) ]$ [\citet{Cressie2011}], and defined as
$2\gamma(\mathbf{s},\mathbf{h}) = E [ Z(\mathbf{s})-
Z(\mathbf{s}+\mathbf{h)}) ]^2$ if the spatial field has a~constant
mean. The function $\gamma(\mathbf{s},\mathbf{h})$ itself is called the
semivariogram. The variogram (or the semivariogram) becomes a function
of $\mathbf{h}$ if the process is stationary, and of $\|\mathbf{h}\|$
if it is also isotropic. For a second-order stationary process, that
is, isotropic, the (theoretical) semivariogram rises from the origin to
the upper asymptote Var$[ Z(\mathbf{s}) ]$ which is called the
\textit{sill} of the semivariogram. The distance at which a certain
fraction of the asymptote is reached is called the \textit{range} of
the semivariogram.

\begin{figure}
\includegraphics{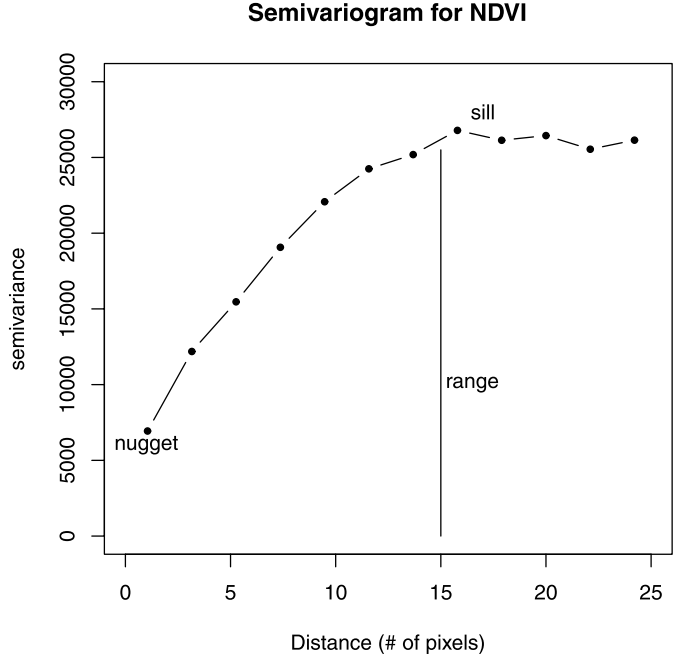}
\caption{The semivariogram plot represents the spatial correlation of
the averaged NDVI values between sites that are 0 to 25 pixels apart.}\label{fig3}
\end{figure}

Given data $Z(\mathbf{s}_i)$, $i=1, \ldots, n$, on $Z(\mathbf{s})$, the
empirical semivariogram is given by $\hat\gamma(h) = \frac{1}{|N_h|}
\sum_{(i,j) \in N_h} [Z(\mathbf{s}_i) - Z(\mathbf{s}_j)]^2$, where
$N_h$ is the set of pairs of observation at locations $\mathbf{s}_i$
and $\mathbf{s}_j$ such that $\|\mathbf{s}_i - \mathbf{s}_j \| = h$,
and $|N_h|$ is the cardinality of this set [\citet{Cressie2011}].
The range can be estimated by plotting the empirical semivariogram
against $h$. When extrapolated to zero distance, the empirical
semivariogram reaches a nonzero value, called a \textit{nugget}, caused
from sampling error resulting in dissimilar values for samples at
locations close to each other. See Figure~\ref{fig3} for illustration
in the NDVI application.

In our present application, we implicitly assume that the dependency
among the NDVI values is localized. In fact, as the first law of
geography states in \citet{Tobler1970}, ``everything is related to
everything else, but near things are more related than distant
things.'' The range estimated from the empirical semivariogram based on
the NDVI values can provide an idea of the size of neighborhoods or
subregions capturing this local dependency. For instance, let $\hat r$
be the estimated range. Then, one may consider creating subregions of a
size $D \times D$ grid box of land, where $D$ is the number of pixels
greater than or equal to $\hat r$. The local dependency will be mostly
concentrated within these subregions. Although spatial autocorrelations
would still be present to some extent among the NDVI values for pixels
in both sides of the boundaries, we will ignore them for the time being
in order to theoretically develop our proposed procedures in the next
subsection.

%

\subsection{The proposed procedures}\label{sec3.3}

Suppose that the East African region is divided into subregions of a
size $D \times D$ grid box of land, where $D$ is the number of pixels
which is determined by the range of the semivariogram plot, as
described above. Each pixel, an 8~km${}\times{}$8~km grid box of land,
will be referred to as a ``location.'' Note that some locations in a
subregion may be missing in the sense of containing only water and
hence producing no NDVI observations. However, we will only consider
the subregions with at least one nonmissing location. Let $m$ be the
number of such subregions and $n_i$ be the number of locations in the
$i$th subregion.

We use a two-sided monotonic trend test for each of the four seasons in
a location using the Brillinger test as described in
Section~\ref{sec2}. With $\beta_{ijk}$ being the monotonic trend
parameter as defined in the Brillinger test for the $i$th subregion,
$j$th location and $k$th season, where $i = 1, \ldots, m$, $j =
1,\ldots, n_i$, and $k = 1, 2, 3, 4$, let $T_{ijk}$ and $P_{ijk}$ be,
respectively, the test statistic and the corresponding $p$-value for
testing the null hypothesis $H_{ijk}\dvtx  \beta_{ijk}=0$ against its
two-sided alternative. Using a Bonferroni correction over the seasons,
we define the so-called combined $p$-value for the $j$th location in
the $i$th subregion as $P_{ij} = 4 \min_{1 \le k \le4} P_{ijk}$. The
combined $p$-value for the $i$th subregion is defined as $P_i = n_i
\min_{1 \le j \le n_i} P_{ij}$, which is a Bonferroni correction over
the seasons and locations. With $H_{ijk}$ representing the null
hypothesis corresponding to $P_{ijk}$, we consider $H_{i} =
\bigcap_{j=1}^{n_i} \bigcap_{k=1}^4 H_{ijk}$ as the null hypothesis
corresponding to $i$th subregion, and $H_{ij} = \bigcap_{k}^4 H_{ijk}$
as the null hypothesis corresponding to the $j$th location in the $i$th
subregion.

We propose a procedure that tests the $H_{ijk}$'s against their
respective two-sided alternatives and detects the directions of the
alternatives for the rejected $H_{ijk}$'s. It operates in three stages,
by testing $H_i$, $i=1, \ldots, m$, at the first stage; $H_{ij} =
\bigcap_{k}^4 H_{ijk}$, $j=1, \dots, n_i$, for each $i$ such that $H_i$
is rejected, at the second stage; and $H_{ijk}$, $k=1, \ldots, 4$, for
each $(i,j)$ such that $H_{ij}$ is rejected, at the third stage. More
specifically, our first procedure is defined as follows:

\begin{pro}[(Three-stage directional BH)]\label{pro1}
\begin{longlist}[\textit{Stage} 3.]
\item[\textit{Stage} 1.] Apply the BH method to test $H_i$, $i=1,
\ldots, m$, based on their respective $p$-values $P_1, \ldots,
P_m$ as follows: consider the (increasingly) ordered versions
of the $P_i$'s, $P_{(1)} \le\cdots\le P_{(m)}$, find $S= \max\{
i\dvtx P_{(i)} \le i \alpha/m \}$, and reject the $H_i$'s for
which the $p$-values are less than or equal to $P_{(S)}$,
provided this maximum exists, otherwise, accept all $H_i$.

\item[\textit{Stage} 2.] For every $i$ such that $H_i$ is rejected at
stage~1, consider testing $H_{ij}$, $j=1, \ldots, n_i$, based
on their respective $p$-values $P_{ij}$, $j=1, \ldots, n_i$, as
follows: reject $H_{ij}$ if $P_{ij} \le S \alpha/\{m n_i\}$.

\item[\textit{Stage} 3.] For every $(i,j)$ such that $H_{ij}$ is
rejected at stage~2, first consider testing $H_{ijk}$, $k=1,
\ldots, 4$, based on their respective $p$-values $P_{ijk}$,
$k=1, \ldots, 4$, as follows: reject $H_{ijk}$ if $P_{ijk} \le
S\alpha/\{4mn_i\}$; then, for each rejected $H_{ijk}$, decide
the direction of the monotonic trend to be the same as that of
sign($T_{ijk}$).
\end{longlist}
\end{pro}

The first two stages in Procedure~\ref{pro1} identify the locations
with significant vegetation changes, while the third stage allows one
to make a more detailed analysis for each significant location by
specifying the seasons that contribute to those changes as well as the
directions in which these changes have occurred.

%
\begin{teo}\label{th3.1}
The three-stage directional BH procedure controls the mdFDR at level
$\alpha$ if the subregions are independent.
\end{teo}

A proof of this theorem is given in an \hyperref[app]{Appendix}.

We should point out that our assumption of dependence within, but not
between, the subregions is made only to provide a theoretical framework
for the development of our procedure in Theorem~\ref{th3.1}, even
though, as said above, there is some dependence among the subregions.
It is important to verify that this procedure can continue to control
the mFDR under a certain type of positive dependence condition among
the subregions, like the one that would be similar to the positive
regression dependence on the subset (PRDS) condition
[\citet{BenjaminiYekutieli2001}, \citet{Sarkar2002}] in the
present context. We will do that numerically, since it seems difficult
to do theoretically.

It is seen when proving the above theorem that the mdFDR of
Procedure~\ref{pro1} is $\le\frac{\alpha}{m} \sum_{i=1}^m \frac{1+
\pi_{i0}}{2}$, where $\pi_{i0}$ is the proportion of true $H_{ijk}$'s
(out of the $4 n_i$ null hypotheses) in the $i$th subregion. If
$\pi_{i0}$ were known, one would have used $\frac{1 + \pi_{i0}}{2}
P_{ijk}$, instead of $P_{ijk}$, in Procedure~\ref{pro1}, to get a
tighter control over the mdFDR at $\alpha$. In reality, when $\pi_{i0}$
is unknown, one can consider estimating it from the data. With that in
mind, we propose our next procedure as an adaptive version of
Procedure~\ref{pro1} by estimating $\pi_{i0}$ using a
\citet{StoreyTaylorSiegmund2004} type estimate.



\begin{pro}[(Adaptive three-stage directional BH)]\label{pro2}
Consider Procedure~\ref{pro1} with $P_{ijk}$ replaced by $\frac{1+
\hat\pi_{i0}}{2} P_{ijk}$, where
%
%
\begin{equation}
\hat{\pi}_{i0} = \min \biggl\{ \frac{
\sum_{j=1}^{n_i} \sum_{k=1}^4 I(P_{ijk} > \lambda) +
1}{4n_i(1-\lambda)}, 1 \biggr\}
\end{equation}
for any $\lambda\in (0,1)$. Typically $\lambda$ is chosen to be 0.5.
\end{pro}

It is important to note how each $P_{ijk}$ is being adjusted in this
adaptive test based on the information shared by the other $p$-values
in each subregion. If $\sum_{j=1}^{n_i} \sum_{k=1}^4 I(P_{ijk} >
\lambda)$ gets larger (or smaller), indicating more (or less)
nonsignificant (or significant) $p$-values in the $i$th subregion, then
$P_{ijk}$ moves further away from (or gets shrunk toward) zero, making
it more likely to be nonsignificant (or significant) also.

\section{Simulation studies}\label{sec4}

We ran a number of simulation studies to examine the mdFDR control
property and the power of our proposed procedures. Keep in mind that
the proposed procedures were developed assuming arbitrary dependence
among locations within each subregion ($j = 1, \ldots, n_i$) and among
the four seasons at each location ($k = 1, \ldots, 4$). To account for
the correlation between pixels, we assume spatial stationarity across
the region. We also assume isotropic spatial autocorrelation, which
means that the process causing the spatial autocorrelation acts in the
same way in all directions. Isotropic correlations depend only on the
distance $d= \|\mathbf{s}_j - \mathbf{s}_{j^{\prime}} \| $ between
locations $j$ and $j^{\prime}$, but not on the direction.

Some frequently used isotropic covariance functions are the exponential
model ($C(d)=\sigma^2 \exp(-d/\theta)$), the Gaussian model
($C(d)=\sigma^2 \exp(-d/\theta)^2$) and the spherical model
($C(d)=\sigma^2 ( 1+\frac{d}{2\theta} ) (1 - \frac{d}{\theta} )^2_+$),
where $\theta$ is a scaling factor which is related to the range of the
variogram. For bounded variograms that reach the sill asymptotically
(e.g., exponential model), in practice, $\theta$ is taken to be the
distance where the model reaches $95\%$ of the sill (also called the
practical range).

In this simulation study, we selected the exponential correlation model
to simulate spatial dependence between pixels within a subregion. More
specifically, we did the simulation studies under the following
dependence scenario.

The $p$-values arise from test statistics $X_{ijk} \sim N(Z_{ijk} \mu,
\bolds{\Sigma})$, $i=1, \ldots, m$, $j=1, \ldots, n_i$, $k=1, \ldots,
4$, where $Z_{ijk}$ are random signals that are i.i.d. Bernoulli
($1-\pi _0$). The covariance matrix, $\bolds{\Sigma}$, includes
correlations on three levels: between season correlation, between pixel
correlation, and between subregion correlation. More specifically, let
$\bolds{\Gamma}_1= (( C(d_{jj'})=\sigma^2 \exp(-d_{jj'}/\theta) )) $,
where $d_{jj'}$ is the distance between location $j$ and $j'$,
$\sigma^2 = 1$, and $\theta$ is estimated by the size of the subregion,
$\bolds{\Gamma}_2=(1- \rho_1) \mathbf{I}_4 + \rho_1 \mathbf{1}_4
\mathbf {1}'_4$, where $-\frac{1}{3} < \rho_1 < 1$, and $\bolds{\Gamma}
_3=(1- \rho_2) \mathbf{I}_m + \rho_2 \mathbf{1}_m \mathbf{1}'_m$, where
$-\frac{1}{m-1} < \rho_2 < 1$, and then define $\bolds{\Sigma} =
\bolds{\Gamma}_1 \otimes\bolds{\Gamma}_2 \otimes\bolds{\Gamma}_3$. In
other words, we assume $\operatorname{Corr} (Z_{ijk},Z_{i'j'k'}) =
\rho_1 \rho_2 \exp(-d_{jj'}/\theta)$, for $i,i'=1, \ldots, m; j,j'=1,
\ldots, n_i;k,k'=1, \ldots, 4$, with $\rho_1=1$ for $k=k'$, and
$\rho_2=1$ for $i=i'$.

We simulated both mdFDR and (average) power, the expected proportion of
correctly rejected among all the false null hypotheses, for both
Procedures \ref{pro1}~and~\ref{pro2} (with $\lambda=0.5$, as often
considered) by choosing $\mu= \pm2, \pm3$ or $\pm5$; $n_i = 9$, 100 or
400; $\pi_0 = 0.9$ or $0.99$; $\rho_1 = -0.3, 0$, $0.4$ or $0.8$; and
$\rho_2 = 0$, 0.2 or 0.5. The reason for selecting a negative value for
$\rho_1$ is that vegetation trends have been noted
[\citet{VrielingdeBeursBrown2008}] to change in different
directions in two successive seasons.

\begin{table}
\tabcolsep=0pt
 \caption{Simulation studies of the mdFDR control and
power of Procedures \protect\ref{pro1} \textup{(P1)}
and~\protect\ref{pro2} \textup{(P2)} and BY under various dependence
scenarios with a nominal error rate of 0.05}\label{tab1}
\begin{tabular*}{\tablewidth}{@{\extracolsep{\fill}}@{}ld{2.1}d{1.1}d{3.0}d{1.11}d{1.11}d{1.11}@{}}
\hline
\textbf{Num.} & \multicolumn{1}{c}{$\bolds{\rho_1}$} & \multicolumn{1}{c}{$\bolds{\rho_2}$}
              & \multicolumn{1}{c}{$\bolds{n_i}$}    & \multicolumn{1}{c}{\textbf{mdFDR$\bolds{{}/{}}$power (P1)}}
              & \multicolumn{1}{c}{\textbf{mdFDR${}\bolds{/}{}$power (P2)}}
              & \multicolumn{1}{c}{\textbf{BY}}
\\
\hline
\phantom{1}1 & -0.3 & 0 & 9 & 0.0248 / 0.3937 & 0.0291 / 0.4175 & 0.0076 / 0.2292\\
\phantom{1}2 & -0.3 & 0.2 & 9 & 0.0302 / 0.2544 & 0.0394 / 0.2827 & 0.0065 /0.1125 \\
\phantom{1}3 & -0.3 & 0.5 & 9 & 0.0338 / 0.0793 & 0.0519 / 0.1019 & 0.0086 /0.0251 \\
\phantom{1}4 & 0 & 0 & 9 & 0.0261 / 0.3915 & 0.0304 / 0.4156 & 0.0090 / 0.2301 \\
\phantom{1}5 & 0 & 0.2 & 9 & 0.0266 / 0.2563 & 0.0352 / 0.2830 & 0.0054 / 0.1151 \\
\phantom{1}6 & 0 & 0.5 & 9 & 0.0398 / 0.0787 & 0.0528 / 0.1020 & 0.0085 / 0.0230 \\
\phantom{1}7 & 0.4 & 0 & 9 & 0.0238 / 0.3899 & 0.0287 / 0.4134 & 0.0074 / 0.2268 \\
\phantom{1}8 & 0.4 & 0.2 & 9 & 0.0281 / 0.2541 & 0.0359 / 0.2824 & 0.0072 / 0.1143\\
\phantom{1}9 & 0.4 & 0.5 & 9 & 0.0370 / 0.0804 & 0.0552 / 0.1035 & 0.0092 / 0.0245\\
10 & 0.8 & 0 & 9 & 0.0245 / 0.3961 & 0.0299 / 0.4196 & 0.0070 / 0.2312\\
11 & 0.8 & 0.2 & 9 & 0.0309 / 0.2551 & 0.0372 / 0.2842 & 0.0085 / 0.1171 \\
12 & 0.8 & 0.5 & 9 & 0.0311 / 0.0799 & 0.0477 / 0.1026 & 0.0079 / 0.0235 \\
13 & -0.3 & 0 & 100 & 0.0062 / 0.2012 & 0.0065 / 0.2080 & 0.0053 / 0.1873 \\
14 & -0.3 & 0.2 & 100 & 0.0098 / 0.1239 & 0.0102 / 0.1292 & 0.0050 / 0.0805 \\
15 & -0.3 & 0.5 & 100 & 0.0227 / 0.0373 & 0.0268 / 0.0427 & 0.0044 / 0.0082 \\
16 & 0 & 0 & 100 & 0.0061 / 0.2019 & 0.0064 / 0.2088 & 0.0051 / 0.1884\\
17 & 0 & 0.2 & 100 & 0.0092 / 0.1245 & 0.0098 / 0.1296 & 0.0053 / 0.0818 \\
18 & 0 & 0.5 & 100 & 0.0237 / 0.0377 & 0.0260 / 0.0430 & 0.0050 /0.0082 \\
19 & 0.4 & 0 & 100 & 0.0060 / 0.2010 & 0.0064 / 0.2078 & 0.0052 /0.1870 \\
20 & 0.4 & 0.2 & 100 & 0.0092 / 0.1242 & 0.0097 / 0.1293 & 0.0051 /0.0808 \\
21 & 0.4 & 0.5 & 100 & 0.0237 / 0.0379 & 0.0273 / 0.0432 & 0.0059 /0.0082 \\
22 & 0.8 & 0 & 100 & 0.0056 / 0.2014 & 0.0061 / 0.2083 & 0.0049 /0.1877 \\
23 & 0.8 & 0.2 & 100 & 0.0091 / 0.1237 & 0.0098 / 0.1288 & 0.0050 /0.0810 \\
24 & 0.8 & 0.5 & 100 & 0.0241 / 0.0381 & 0.0263 / 0.0433 & 0.0059 /0.0084 \\
25 & -0.3 & 0 & 400 & 0.0026 / 0.1214 & 0.0027 / 0.1260 & 0.0046 /0.1717 \\
26 & -0.3 & 0.2 & 400 & 0.0042 / 0.0690 & 0.0044 / 0.0720 & 0.0043 /0.0709 \\
27 & -0.3 & 0.5 & 400 & 0.0147 / 0.0199 & 0.0155 / 0.0213 & 0.0046 /0.0050 \\
28 & 0 & 0 & 400 & 0.0024 / 0.1220 & 0.0026 / 0.1265 & 0.0045 / 0.1724\\
29 & 0 & 0.2 & 400 & 0.0042 / 0.0695 & 0.0044 / 0.0725 & 0.0044 /0.0717 \\
30 & 0 & 0.5 & 400 & 0.0138 / 0.0201 & 0.0146 / 0.0214 & 0.0034 /0.0052 \\
31 & 0.4 & 0 & 400 & 0.0025 / 0.1218 & 0.0026 / 0.1264 & 0.0044 /0.1723 \\
32 & 0.4 & 0.2 & 400 & 0.0041 / 0.0691 & 0.0044 / 0.0721 & 0.0042 /0.0711 \\
33 & 0.4 & 0.5 & 400 & 0.0140 / 0.0201 & 0.0147 / 0.0216 & 0.0045 /0.0053 \\
34 & 0.8 & 0 & 400 & 0.0024 / 0.1223 & 0.0025 / 0.1268 & 0.0043 /0.1733 \\
35 & 0.8 & 0.2 & 400 & 0.0045 / 0.0692 & 0.0048 / 0.0721 & 0.0047 /0.0712 \\
36 & 0.8 & 0.5 & 400 & 0.0142 / 0.0200 & 0.0152 / 0.0214 & 0.0045 /0.0049\\
\hline
\end{tabular*}
\end{table}

The simulated values were obtained based on 1000 simulation runs using
$\alpha= 0.05$. Table~\ref{tab1} compares Procedures
\ref{pro1}~and~\ref{pro2}, and \citet{BenjaminiYekutieli2001} in
terms of these simulated mdFDR and power at several combinations of the
aforementioned chosen values. It is to be noted that the
Benjamini--Yekutieli procedure (BY for short) is an FDR controlling
procedure for arbitrarily correlated $p$-values. Although Procedures
\ref{pro1}~and~\ref{pro2} are designed to control the mdFDR, it is
worth comparing them to existing procedures that have similar
dependence assumptions, namely, the BY. It would not be fair to compare
to methods such as that of \citet{BenjaminiHochberg1995}, since it
requires independence or positive dependence between all $p$-values,
while Procedures \ref{pro1}~and~\ref{pro2} have more relaxed
assumptions.


As seen from Table~\ref{tab1}, the simulated mdFDR of
Procedure~\ref{pro1} remains stably controlled across all correlation
combinations ($\rho_1 = -0.3$, 0, 0.4, 0.8; $\rho_2 = 0$, 0.2, 0.5).
Procedure~\ref{pro2} can outperform Procedure~\ref{pro1}, in the sense
of having higher power while still maintaining control of the mdFDR at
the desired level $\alpha= 0.05$, with weakly to moderately correlated
data within subregions. However, if the data are moderately to largely
correlated with small group sizes ($n_i =9$), Procedure~\ref{pro2} can
lose control of the mdFDR. Although unfortunate, this is not
surprising, knowing that this type of adaptive procedure considered in
the contexts of FDR or FWER control also becomes unstable with large
correlations among the underlying test statistics. We can also see that
as the subregion size increases ($n_i$ from 9 to 400), the power
decreases. This can, however, be attributed to the fact that a
Bonferroni type combination of $p$-values had to be considered, because
of arbitrary dependence, to define subregion and location specific
$p$-values. In comparison, the BY procedure also seems to maintain
control of the mdFDR, even though it is an FDR procedure. However, the
BY procedure has smaller power in every scenario.

\section{East African NDVI results}\label{sec5}

We use the Brillinger test as described in Section~\ref{sec2} to test
for significant vegetation trend over the years separately for the four
seasons at each location. Specifically, with $\bar{Y}_{ijk,t}$
representing the NDVI average for the $i$th subregion, $j$th location
and $k$th season in the $t$th year, where $i=1, \ldots, m$, $j=1,
\ldots, n_i$, and $k=1, \ldots, 4$, we consider, for each fixed
$(i,j,k)$, the following model:
%
%
\begin{equation}
\bar{Y}_{ijk,t} = S_{ijk}(t) + E_{ijk}(t)\qquad
\mbox{for }t=0,1, \ldots, 24
\end{equation}
and test $H_0\dvtx  S_{ijk}(t)$ is a constant signal vs. $H_1\dvtx
S_{ijk}(t)$ is monotonic in time, using the Brillinger test statistic
and the corresponding approximate $p$-value. A negative significant
test statistic provides evidence of a monotonic decreasing trend, while
a positive significant test statistic suggests an increasing monotonic
trend.

We applied Procedures \ref{pro1}~and~\ref{pro2} (with $\lambda=0.5$ and
$\alpha =0.05$) based on the \mbox{$p$-}values for the above tests to
the region to screen for significant seasonal vegetation changes over
the years and the directions in which these changes are taking place.

As mentioned before, we used the semivariogram plot to determine the
grid size ($D \times D$) for each subregion, where each site's averaged
NDVI value over all years is $Z(\mathbf{s})$. The empirical
semivariogram plot for the NDVI data is shown in Figure~\ref{fig3}. As seen from
this plot, the range of the semivariogram is approximately 15 pixels,
meaning NDVI values for locations with a Euclidean distance greater
than 15 pixels apart are uncorrelated. Thus, it would be appropriate to
choose the group size $D \ge15$.

In particular, each subregion's minimum $p$-value ($P_i$) is used to
represent the corresponding group in the stage~1 BH method in our
procedure. Thus, if the locations of $P_i$ and $P_j$ are at least 15
pixels apart, we can consider according to this semivariogram plot that
the corresponding subregions are independent.

After carefully considering various group sizes of $D \ge15$, we
choose $D=20$, yielding $G=150$ groups, each with $n_i \le400$
locations. Using $D=20$ to group the locations, only 1.37\% of the
group minima were closer than 15 pixels to another group minimum. Thus,
we are satisfied that each group, represented by the group minimum, is
nearly independent from the others.

This region has several bodies of water, including the African Great
Lakes and part of the Indian Ocean, where there is no vegetation. Thus,
the remote sensing pixels covering entirely water will correspond to a
missing location in a subregion's grid, that is, the subregions that
straddle land and water will have $n_i < 400$.

%
\begin{figure}
\includegraphics{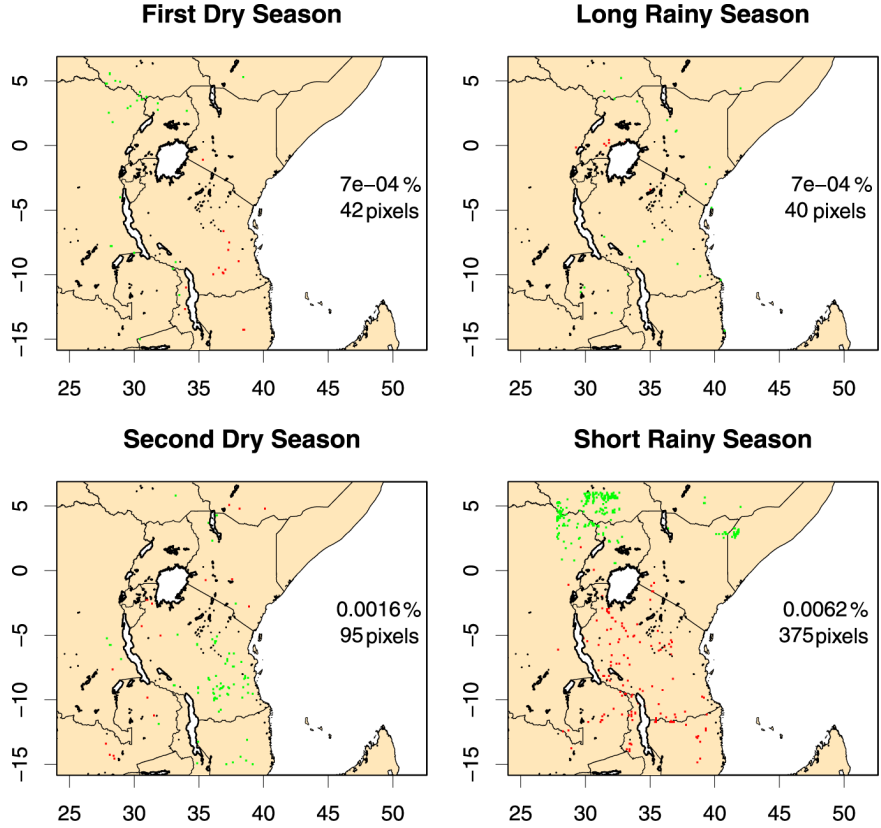}
\caption{The results of applying Procedure~\protect\ref{pro1}, where 42, 40, 95 and 375
of the pixel's $p$-values were found to have significant increasing or
decreasing changes in their respective seasonal NDVI averages. Sites
with a significant seasonal increasing change in vegetation are plotted
in green, significant seasonal negative vegetation change are plotted
in red, and nonsignificant sites are represented by tan.}\label{fig4}
\end{figure}

%
\begin{figure}
\includegraphics{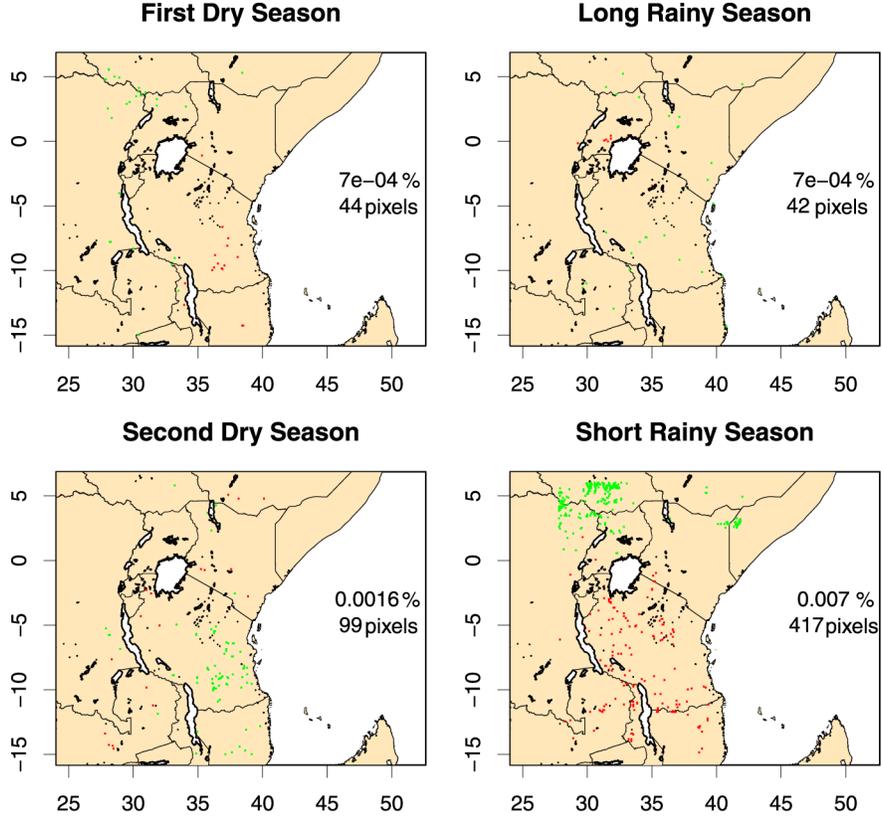}
\caption{The results of applying Procedure~\protect\ref{pro2}. The number of pixels with
significant trend changes are 44, 42, 99 and 417 in their respective
seasonal NDVI averages. Sites with a significant seasonal increasing
change in vegetation are plotted in green, significant seasonal
negative vegetation change are plotted in red, and nonsignificant sites
are represented by tan.}\label{fig5}
\end{figure}

The results of Procedures \ref{pro1}~and~\ref{pro2} for each of the
four seasons are shown in Figures~\ref{fig4} and \ref{fig5},
respectively. Sites with a significant seasonal increasing change in
vegetation are plotted in green. Sites with significant seasonal
negative vegetation change are plotted in red. The nonsignificant sites
are represented by tan. Using Procedure~\ref{pro1}, we detected $42$,
$40$, $95$ and $375$ pixels with significant increasing or decreasing
changes in their respective seasonal NDVI averages (first dry season,
long rainy season, second dry season and short rainy season). The
second dry season has concentrated locations in coastal and central
Tanzania with increasing average NDVI and the short rainy season has
concentrated decreasing vegetation changes directly South of Lake
Victoria, both of which are potentially important findings for land use
managers. Using Procedure~\ref{pro2}, the number of pixels with
significant $p$-values increases to 44, 42, 99 and 417 in their
respective seasonal NDVI averages in generally the same regions as
found from Procedure~\ref{pro1}.

Overall, the results show increasing vegetation trends in the Northern
hemisphere as well as coastal Eastern Tanzania. Decreasing vegetation
trends are mostly concentrated directly South of Lake Victoria. Another
noticeable finding is that the second dry season and short rainy season
(which make up the last 6 months of the calendar year) are the seasons
that contain the majority of the significance. These findings are
consistent with historical evidence and other climate change
investigations done in this region, as described below. However, this
is the first study to reach these findings while maintaining control of
a meaningful level of Types I~and~III errors.
\section{Discussion and concluding remarks}\label{sec6}
The motivation of this paper lies in the fact that the currently
available statistical approach to detecting vegetation changes over
different locations in a region, like in East Africa, developed so far
in the framework of testing multiple hypotheses [e.g., in
\citet{VrielingdeBeursBrown2008}], may be questionable. Current
methodologies have not taken into account the multiplicity by guarding
against an overall measure of false discoveries, directional and
nondirectional, and hence can potentially produce too many falsely
discovered vegetation locations, more than what is statistically
acceptable. It is important to avoid falsely discovered locations in
the present application since the data findings could be used to
relocate East African populations of people, livestock and crops, which
is risky and costly.

Our findings, in terms of the proportion of regions with discovered
vegetation change, are notably a smaller subset of the conclusions from
other studies in this region, including
\citet{ColeDunbarMcClanahanMuthiga2000},
\citet{VrielingdeBeursBrown2008}, \citet{UsongoNagahuedi2008}
and \citet{DuveillerDefournyDescleeMayaux2007}. First to analyze
this data, \citet{VrielingdeBeursBrown2008} used trend tests on
several phenology indicators at every location in the region and found
a substantial proportion of ``significant'' changes in all
indicators---as high as 44\%. However, their conclusions failed to
address any control on the error rate while testing thousands of
hypotheses simultaneously.

Addressing this multiplicity issue is important, since making false
claims about significant vegetation change is costly when it involves
risking the livelihood of entire populations of people and livestock.
Our proposed methods not only address the multiplicity by controlling
an overall measure of combined directional and nondirectional false
discoveries, the mdFDR, but also are developed with the idea of making
them as powerful as possible by adequately capturing spatial dependency
present in the data. Since sites tend to be dependent more locally than
globally, we consider grouping the hypotheses into suitable clusters
before developing these methods to be a way of capturing such local
dependency. The idea of using grouped hypotheses has been successfully
used in \citet{ClementsSarkarGuo2011} in a two-stage format and to
control the FDR. By augmenting this procedure to include directional
errors, we are able to detect important directional vegetation changes
in all four precipitation seasons in the East African region, while
maintaining control of the mdFDR. It is important to point out,
however, that Procedure~\ref{pro1} is one that offers an mdFDR
controlling procedure in the present setting, that is, robust against
spatial dependency but does not explicitly use such dependency
(quantitatively speaking). Its adaptive version, Procedure~\ref{pro2},
attempts to explicitly use such spatial dependency.

To reiterate, controlling an FDR related error rate incorporating
directional errors, like the mdDFR, is the rationale behind proposing
our procedures, since detecting areas with significant increasing or
decreasing vegetation change is one of the primary objectives in the
present application. Our procedure is an augmented version of an FDR
controlling procedure, similar to \citet{BenjaminiYekutieli2005}.
There are other FDR controlling procedures proposed in similar
\mbox{cluster} settings [\citet{BenjaminiHeller2007},
\citet{PacificoVerdinelliGenoveseWasserman2004}]. However, it
remains to be determined if these procedures can be augmented as in
Procedure~\ref{pro1} without losing control over the mdFDR. Then, one
can consider these procedures as relevant competitors of ours and
evaluate the performance of our procedure relative to them.

Utilizing the information gleaned after applying Procedures
\ref{pro1}~and~\ref{pro2}, more sophisticated modeling techniques can
be applied to the locations with seasonal changes. By first detecting
locations of change using multiple testing, we are protected from
investigating too many falsely discovered locations. Specifically,
spatio-temporal modeling and forecasting may be of interest to land use
management to optimize utilization of the land based on projected NDVI.

There are a few areas for improvement to this study. Although the
proposed procedures do not require any dependence assumption for the
$p$-values within each subregion, the theoretical proof demands that
subregions be independent to maintain control of the mdFDR. It would be
interesting to theoretically investigate the performance of the
proposed three-stage directional procedures under more complex
subregion dependence structures. Second, selection of the optimal
subregion size $D$ is debatable, as with any tuning parameter. This
parameter needs to be large enough such that one can reasonably assume
the subregions are independent, yet keeping in mind that, in light of
the simulation studies, larger group sizes yield less powerful
procedures due to the Bonferroni adjustments. The idea of estimating
the range of a variogram is one such way to select the subregion size
$D$.

\begin{appendix}\label{app}
\section*{Appendix: Proof of Theorem~3.1}

\begin{pf}
Let $R$ be the total number of $H_{ijk}$'s that have been rejected, and
$V$~and~$U$, respectively, be the numbers of Types I~and~III errors
that occurred out of these $R$ rejections. Then
\[
\mathrm{mdFDR} = E \biggl( \frac{V+U}{\max\{R,1\}} \biggr) = \mathrm{FDR} +
\mathrm{dFDR},
\]
where $\mathrm{FDR} = E (\frac{V}{\max\{R,1\}} )$ is the FDR, and
$\mathrm{dFDR} = E (\frac{U}{\max\{R,1\}} )$ is the (pure) directional
FDR.

Let us consider using $H_{ijk}$ also as an indicator variable with
$H_{ijk}=0$ (or $1$), indicating that the null hypothesis $H_{ijk}\dvtx
\beta_{ijk} = 0$ is true (or false). Then,
\[
V = \sum_{i=1}^m \sum
_{j=1}^{n_i} \sum_{k=1}^4
I \bigl(H_{ijk}=0, P_{ijk} \le S \alpha/\{4mn_i\}
\bigr),
\]
where $S$ is the number of significant subregions in the first stage of
the procedure. Hence,
%
%
\begin{eqnarray}\label{A.1}
\mathrm{FDR} & = & E \biggl(\frac{V}{\max\{R, 1 \}} \biggr)\nonumber
\\
& = & \sum_{i=1}^m \sum
_{j=1}^{n_i} \sum_{k=1}^4
E \biggl( \frac
{ I(H_{ijk}=0, P_{ijk} \le S \alpha/\{4mn_i\})}{\max\{R,1\}} \biggr)
\\
& \le& \sum_{i=1}^m \sum
_{j=1}^{n_i} \sum_{k=1}^4
I(H_{ijk}=0) E \biggl( \frac{I(P_{ijk} \le S \alpha/\{4mn_i\})}{\max\{S,1\}} \biggr),\nonumber
\end{eqnarray}
since $R \ge S$ [borrowing the idea from \citet{GuoSarkar2012}].
Let $S^{(-i)}$ be the number of significant subregions that would have
been obtained if we had completely ignored the $i$th subregion and
applied the first-stage BH method to the rest of the $m-1$ subregion
$p$-values using the critical values $i\alpha/m$, $i=2, \ldots, m$.
Then, it can be shown that
%
%
\begin{eqnarray}\label{A.2}
\qquad\frac{I (P_{ijk} \le S \alpha/\{4mn_i\} )}{\max\{S,1\}}&=& \sum_{s=1}^m\frac{I (P_{ijk} \le s \alpha/\{4mn_i\}, S=s)}{s}
\nonumber\\[-8pt]\\[-8pt]
&=& \sum_{s=1}^m \frac{I (P_{ijk} \le s \alpha/\{4mn_i\}, S^{(-i)} = s-1 )}{s}.\nonumber
\end{eqnarray}
Since the subregions are assumed independent, taking expectation in
(\ref{A.2}) and using that in (\ref{A.1}), we see that
%
%
\begin{eqnarray}
\mathrm{FDR} & \le& \sum_{i=1}^m \sum
_{j=1}^{n_i} \sum_{k=1}^4
I(H_{ijk}=0) \sum_{s=1}^m
\frac{1}{s} \frac{s\alpha}{4mn_i} \operatorname{Pr} \bigl( S^{(-i)} = s-1 \bigr)
\nonumber\\[-8pt]\\[-8pt]
&=& \alpha\sum_{i=1}^m \frac{1}{4mn_i}\sum_{j=1}^{n_i} \sum
_{k=1}^4 I(H_{ijk}=0) =\frac{\alpha}{m} \sum_{i=1}^m\pi_{i0},\nonumber
\end{eqnarray}
where $\pi_{i0}$ is the proportion of true null hypotheses among the
total $4n_i$ null hypotheses in the $i$th subregion.

We now work with the dFDR. With $\delta_{ijk}= \sign(\beta_{ijk})$
representing the true sign of the Brillinger's monotonic trend
parameter $\beta_{ijk}$, $U$ can be expressed as follows:
\[
U = \sum_{i=1}^m \sum
_{j=1}^{n_i} \sum_{k=1}^4
I \bigl(H_{ijk} = 1, P_{ijk} \le S \alpha/\{4mn_i
\}, T_{ijk} \delta_{ijk} < 0 \bigr)
\]
from which we first have
\begin{eqnarray*}
\mathrm{dFDR} & = & E \biggl(\frac{U}{\max\{R, 1 \}} \biggr)
\\
& = & \sum_{i=1}^m \sum
_{j=1}^{n_i} \sum_{k=1}^4
I(H_{ijk}=1) E \biggl( \frac{ I(P_{ijk} \le S \alpha/\{4mn_i\}, T_{ijk}
\delta_{ijk} < 0 )}{\max\{R,1\}} \biggr).
\end{eqnarray*}
Making arguments similar to those used for the FDR, we then have
%
%
\begin{eqnarray}\label{A.4}
\qquad\quad \mathrm{dFDR} & \le& \sum_{i=1}^m \sum
_{j=1}^{n_i} \sum_{k=1}^4
I(H_{ijk}=1)
\nonumber\\[-8pt]\\[-8pt]
&&{} \times\sum_{s=1}^m \frac{1}{s}
\operatorname{Pr} \bigl( P_{ijk} \le s\alpha/\{4mn_i\}, T_{ijk}
\delta_{ijk} < 0 \bigr) \operatorname{Pr} \bigl( S^{(-i)} = s-1 \bigr).\nonumber
\end{eqnarray}
Notice that $P_{ijk} = 2 [1- \Phi (|T_{ijk}| ) ]$, where $\Phi$ is the
cumulative distribution function of the standard normal. Therefore,
assuming without any loss of generality that $\beta_{ijk} > 0$ when
$H_{ijk} =1$, we have, for such $H_{ijk}$,
%
%
\begin{eqnarray}\label{A.5}
&& \operatorname{Pr} \bigl( P_{ijk} \le s\alpha/\{4mn_i\},
T_{ijk} \delta _{ijk} < 0 \bigr)
\nonumber
\\
&&\qquad = \operatorname{Pr}_{\beta_{ijk} >0} \biggl( |T_{ijk}| \ge F^{-1}
\biggl(1 - \frac{s\alpha}{8mn_i} \biggr), T_{ijk} < 0 \biggr)\nonumber
\\
&&\qquad = \operatorname{Pr}_{\beta_{ijk} >0} \biggl( T_{ijk} \le - F^{-1}
\biggl(1 - \frac{s\alpha}{8mn_i} \biggr) \biggr)
\\
&&\qquad \le \operatorname{Pr}_{\beta_{ijk} =0} \biggl( T_{ijk} \le - F^{-1}
\biggl(1 - \frac{s\alpha}{8mn_i} \biggr) \biggr)\nonumber
\\
&&\qquad = \frac{s\alpha}{8mn_i}.\nonumber
\end{eqnarray}
The last inequality follows from the fact that, when $H_{ijk}=1$, the
distribution of~$T_{ijk}$ is stochastically increasing in $\beta
_{ijk}$. Using (\ref{A.5}) in (\ref{A.4}), we see that
%
%
\begin{eqnarray}
\mathrm{dFDR} & \le& \frac{\alpha}{2m} \sum_{i=1}^m
\frac{1}{4n_i} \sum_{j=1}^{n_i} \sum
_{k=1}^4 I(H_{ijk}=1) =
\frac{\alpha}{2m} \sum_{i=1}^m\pi_{i1},
\end{eqnarray}
where $\pi_{i1}$ is the proportion of false null hypotheses among the
total $4n_i$ null hypotheses in the $i$th subregion.

Thus, we finally have
%
%
\begin{equation}
\mathrm{mdFDR} \le\frac{\alpha}{m} \sum_{i=1}^m
\biggl( \pi _{i0} + \frac{1}{2} \pi_{i1} \biggr) = \frac{\alpha}{m}
\sum_{i=1}^m \biggl( \frac{1+\pi_{i0}}{2} \biggr) \le\alpha,
\end{equation}
proving the desired result.
\end{pf}
\end{appendix}

\section*{Acknowledgments}
The NDVI data set was collected as part of a Michigan State University
research project, namely, the ``Dynamic Interactions among People,
Livestock, and Savanna Ecosystems under Climate Change'' project
(funded by the National Science Foundation Biocomplexity of Coupled
Human and Natural Systems Program, Award No. BCS/CNH 0709671). We thank
the anonymous referees for their constructive comments which have
helped to improve the quality of the paper.


%

\printaddresses


\begin{thebibliography}{25}
\bibitem[\protect\citeauthoryear{Abelson and Tukey}{1963}]{AbelsonTukey1963}
%
\begin{barticle}[mr]
\bauthor{\bsnm{Abelson},~\bfnm{Robert~P.}\binits{R.~P.}} \AND
\bauthor{\bsnm{Tukey},~\bfnm{John~W.}\binits{J.~W.}}
(\byear{1963}).
\btitle{Efficient utilization of non-numerical information in quantitative
analysis: {G}eneral theory and the case of simple order}.
\bjournal{Ann. Math. Statist.}
\bvolume{34}
\bpages{1347--1369}.
\bid{issn={0003-4851}, mr={0156411}}
\bptok{imsref}%
\end{barticle}
%
\endbibitem

\bibitem[\protect\citeauthoryear{Benjamini and
Heller}{2007}]{BenjaminiHeller2007}
%
\begin{barticle}[mr]
\bauthor{\bsnm{Benjamini},~\bfnm{Yoav}\binits{Y.}} \AND
\bauthor{\bsnm{Heller},~\bfnm{Ruth}\binits{R.}}
(\byear{2007}).
\btitle{False discovery rates for spatial signals}.
\bjournal{J. Amer. Statist. Assoc.}
\bvolume{102}
\bpages{1272--1281}.
\bid{doi={10.1198/016214507000000941}, issn={0162-1459}, mr={2412549}}
\bptok{imsref}%
\end{barticle}
%
\endbibitem

\bibitem[\protect\citeauthoryear{Benjamini and
Hochberg}{1995}]{BenjaminiHochberg1995}
%
\begin{barticle}[mr]
\bauthor{\bsnm{Benjamini},~\bfnm{Yoav}\binits{Y.}} \AND
\bauthor{\bsnm{Hochberg},~\bfnm{Yosef}\binits{Y.}}
(\byear{1995}).
\btitle{Controlling the false discovery rate: A practical and powerful approach
to multiple testing}.
\bjournal{J. Roy. Statist. Soc. Ser. B}
\bvolume{57}
\bpages{289--300}.
\bid{issn={0035-9246}, mr={1325392}}
\bptok{imsref}%
\end{barticle}
%
\endbibitem

\bibitem[\protect\citeauthoryear{Benjamini and
Yekutieli}{2001}]{BenjaminiYekutieli2001}
%
\begin{barticle}[mr]
\bauthor{\bsnm{Benjamini},~\bfnm{Yoav}\binits{Y.}} \AND
\bauthor{\bsnm{Yekutieli},~\bfnm{Daniel}\binits{D.}}
(\byear{2001}).
\btitle{The control of the false discovery rate in multiple testing under
dependency}.
\bjournal{Ann. Statist.}
\bvolume{29}
\bpages{1165--1188}.
\bid{doi={10.1214/aos/1013699998}, issn={0090-5364}, mr={1869245}}
\bptok{imsref}%
\end{barticle}
%
\endbibitem

\bibitem[\protect\citeauthoryear{Benjamini and
Yekutieli}{2005}]{BenjaminiYekutieli2005}
%
\begin{barticle}[mr]
\bauthor{\bsnm{Benjamini},~\bfnm{Yoav}\binits{Y.}} \AND
\bauthor{\bsnm{Yekutieli},~\bfnm{Daniel}\binits{D.}}
(\byear{2005}).
\btitle{False discovery rate-adjusted multiple confidence intervals for
selected parameters}.
\bjournal{J. Amer. Statist. Assoc.}
\bvolume{100}
\bpages{71--93}.
\bid{doi={10.1198/016214504000001907}, issn={0162-1459}, mr={2156820}}
\bptok{imsref}%
\end{barticle}
%
\endbibitem

\bibitem[\protect\citeauthoryear{Brillinger}{1989}]{Brillinger1989}
%
\begin{barticle}[mr]
\bauthor{\bsnm{Brillinger},~\bfnm{David~R.}\binits{D.~R.}}
(\byear{1989}).
\btitle{Consistent detection of a monotonic trend superposed on a stationary
time series}.
\bjournal{Biometrika}
\bvolume{76}
\bpages{23--30}.
\bid{doi={10.1093/biomet/76.1.23}, issn={0006-3444}, mr={0991419}}
\bptok{imsref}%
\end{barticle}
%
\endbibitem

\bibitem[\protect\citeauthoryear{Chen, Jonsson and
Tamura}{2004}]{ChenJonssonTamura2004}
%
\begin{barticle}[author]
\bauthor{\bsnm{Chen},~\bfnm{J.}\binits{J.}},
\bauthor{\bsnm{Jonsson},~\bfnm{P.}\binits{P.}} \AND
\bauthor{\bsnm{Tamura},~\bfnm{M.}\binits{M.}}
(\byear{2004}).
\btitle{A simple method for reconstructing a high-quality NDVI
time-series data set based on the Savitzky--Golay filter}.
\bjournal{Remote Sensing of Environment}
\bvolume{91}
\bpages{332--344}.
\bptok{imsref}%
\end{barticle}
%
\endbibitem

\bibitem[\protect\citeauthoryear{Clements, Sarkar and Guo}{2011}]{ClementsSarkarGuo2011}
%
\begin{binproceedings}[author]
\bauthor{\bsnm{Clements},~\bfnm{N.}\binits{N.}},
\bauthor{\bsnm{Sarkar},~\bfnm{S.~K.}\binits{S.~K.}} \AND
\bauthor{\bsnm{Guo},~\bfnm{W.}\binits{W.}}
(\byear{2011}).
\btitle{Astronomical transient detection controlling the false discovery rate}.
In \bbooktitle{Statistical Challenges in Modern Astronomy~V}
(\beditor{\binits{E.~D.}~\bsnm{Feigelson}} \AND
\beditor{\binits{G.~J.}~\bsnm{Babu}}, eds.)
\bpages{383--396}.
\bpublisher{Springer}, \blocation{New York}.
\bptok{imsref}%
\end{binproceedings}
%
\endbibitem




\bibitem[\protect\citeauthoryear{Cole
et~al.}{2000}]{ColeDunbarMcClanahanMuthiga2000}
%
\begin{barticle}[author]
\bauthor{\bsnm{Cole},~\bfnm{J.~E.}\binits{J.~E.}},
\bauthor{\bsnm{Dunbar},~\bfnm{R.~B.}\binits{R.~B.}},
\bauthor{\bsnm{McClanahan},~\bfnm{T.~R.}\binits{T.~R.}} \AND
\bauthor{\bsnm{Muthiga},~\bfnm{N.~A.}\binits{N.~A.}}
(\byear{2000}).
\btitle{Tropical pacific forcing of decadal SST variability in the
western Indian Ocean over the past two centuries}.
\bjournal{Science}
\bvolume{287}
\bpages{617--619}.
\bptok{imsref}%
\end{barticle}
%
\endbibitem

\bibitem[\protect\citeauthoryear{Cressie and Wikle}{2011}]{Cressie2011}
%
\begin{bbook}[mr]
\bauthor{\bsnm{Cressie},~\bfnm{Noel}\binits{N.}} \AND
\bauthor{\bsnm{Wikle},~\bfnm{Christopher~K.}\binits{C.~K.}}
(\byear{2011}).
\btitle{Statistics for Spatio-Temporal Data}.
\bpublisher{Wiley}, \blocation{Hoboken, NJ}.
\bid{mr={2848400}}
\bptok{imsref}%
\end{bbook}
%
\endbibitem

\bibitem[\protect\citeauthoryear{Curran}{1980}]{Curran1980}
%
\begin{barticle}[author]
\bauthor{\bsnm{Curran},~\bfnm{P.~J.}\binits{P.~J.}}
(\byear{1980}).
\btitle{Multispectral remote sensing of vegetation amount}.
\bjournal{Progress in Physical Geography}
\bvolume{4}
\bpages{315--341}.
\bptok{imsref}%
\end{barticle}
%
\endbibitem

\bibitem[\protect\citeauthoryear{Duveiller
et~al.}{2007}]{DuveillerDefournyDescleeMayaux2007}
%
\begin{barticle}[author]
\bauthor{\bsnm{Duveiller},~\bfnm{G.}\binits{G.}},
\bauthor{\bsnm{Defourny},~\bfnm{P.}\binits{P.}},
\bauthor{\bsnm{Desclee},~\bfnm{B.}\binits{B.}} \AND
\bauthor{\bsnm{Mayaux},~\bfnm{P.}\binits{P.}}
(\byear{2007}).
\btitle{Deforestation in Central Africa: Estimates at regional,
national and
landscape levels by advanced processing of systematically-disturbed
Landsat extracts}.
\bjournal{Remote Sensing of Environment}
\bvolume{112}
\bpages{1969--1981}.
\bptok{imsref}%
\end{barticle}
%
\endbibitem

\bibitem[\protect\citeauthoryear{Foody}{2003}]{Foody2003}
%
\begin{barticle}[author]
\bauthor{\bsnm{Foody},~\bfnm{G.~M.}\binits{G.~M.}}
(\byear{2003}).
\btitle{Geographical weighting as a further refinement to regression modeling:
An example focused on the NDVI--rainfall relationship}.
\bjournal{Remote Sensing of Environment}
\bvolume{88}
\bpages{283--293}.
\bptok{imsref}%
\end{barticle}
%
\endbibitem

\bibitem[\protect\citeauthoryear{Guo and Sarkar}{2012}]{GuoSarkar2012}
%
\begin{bmisc}[author]
\bauthor{\bsnm{Guo},~\bfnm{W.}\binits{W.}} \AND
\bauthor{\bsnm{Sarkar},~\bfnm{S.}\binits{S.}}
(\byear{2012}).
\bhowpublished{Adaptive controls of the FWER and FDR under block dependence.
Unpublished manuscript. Available at
\texttt{%
\href{http://web.njit.edu/\textasciitilde wguo/research.html}{http://}
\href{http://web.njit.edu/\textasciitilde wguo/research.html}{web.njit.edu/\textasciitilde
wguo/research.html}}.}
\bptok{imsref}%
\end{bmisc}
%
\endbibitem

\bibitem[\protect\citeauthoryear{Hayes and Sader}{2001}]{HayesSader2001}
%
\begin{barticle}[author]
\bauthor{\bsnm{Hayes},~\bfnm{D.~J.}\binits{D.~J.}} \AND
\bauthor{\bsnm{Sader},~\bfnm{S.~A.}\binits{S.~A.}}
(\byear{2001}).
\btitle{Comparison of change-detection techniques for monitoring tropical
forest clearing and vegetation regrowth in a time series}.
\bjournal{Photogrammetric Engineering and Remote Sensing}
\bvolume{67}
\bpages{1067--1075}.
\bptok{imsref}%
\end{barticle}
%
\endbibitem

\bibitem[\protect\citeauthoryear{Jackson, Slater and
Pinter}{1983}]{JacksonSlaterPinter1983}
%
\begin{barticle}[author]
\bauthor{\bsnm{Jackson},~\bfnm{R.~D.}\binits{R.~D.}},
\bauthor{\bsnm{Slater},~\bfnm{P.~N.}\binits{P.~N.}} \AND
\bauthor{\bsnm{Pinter},~\bfnm{P.~J.}\binits{P.~J.}}
(\byear{1983}).
\btitle{Discrimination of growth and water stress in wheat by various
vegetation indices through clear and turbid atmospheres}.
\bjournal{Remote Sensing of Environment}
\bvolume{13}
\bpages{187--208}.
\bptok{imsref}%
\end{barticle}
%
\endbibitem

\bibitem[\protect\citeauthoryear{OCHA}{2011}]{OCHA2011}
%
\begin{bmisc}[author]
\bauthor{\bsnm{OCHA}}
(\byear{2011}).
\bhowpublished{Eastern Africa drought humanitarian report No.~3.
\textit{OCHA, UN Office for the Coordination of Humanitarian Affairs}}
reliefweb.int.
\bptok{imsref}%
\end{bmisc}
%
\endbibitem

\bibitem[\protect\citeauthoryear{Pacifico
et~al.}{2004}]{PacificoVerdinelliGenoveseWasserman2004}
%
\begin{barticle}[mr]
\bauthor{\bsnm{Pacifico},~\bfnm{M.~P.}\binits{M.~P.}},
\bauthor{\bsnm{Genovese},~\bfnm{C.}\binits{C.}},
\bauthor{\bsnm{Verdinelli},~\bfnm{I.}\binits{I.}} \AND
\bauthor{\bsnm{Wasserman},~\bfnm{L.}\binits{L.}}
(\byear{2004}).
\btitle{False discovery control for random fields}.
\bjournal{J. Amer. Statist. Assoc.}
\bvolume{99}
\bpages{1002--1014}.
\bid{doi={10.1198/0162145000001655}, issn={0162-1459}, mr={2109490}}
\bptok{imsref}%
\end{barticle}
%
\endbibitem

\bibitem[\protect\citeauthoryear{Sarkar}{2002}]{Sarkar2002}
%
\begin{barticle}[mr]
\bauthor{\bsnm{Sarkar},~\bfnm{Sanat~K.}\binits{S.~K.}}
(\byear{2002}).
\btitle{Some results on false discovery rate in stepwise multiple testing
procedures}.
\bjournal{Ann. Statist.}
\bvolume{30}
\bpages{239--257}.
\bid{doi={10.1214/aos/1015362192}, issn={0090-5364}, mr={1892663}}
\bptok{imsref}%
\end{barticle}
%
\endbibitem

\bibitem[\protect\citeauthoryear{Storey, Taylor and
Siegmund}{2004}]{StoreyTaylorSiegmund2004}
%
\begin{barticle}[mr]
\bauthor{\bsnm{Storey},~\bfnm{John~D.}\binits{J.~D.}},
\bauthor{\bsnm{Taylor},~\bfnm{Jonathan~E.}\binits{J.~E.}} \AND
\bauthor{\bsnm{Siegmund},~\bfnm{David}\binits{D.}}
(\byear{2004}).
\btitle{Strong control, conservative point estimation and simultaneous
conservative consistency of false discovery rates: A unified approach}.
\bjournal{J. R. Stat. Soc. Ser. B Stat. Methodol.}
\bvolume{66}
\bpages{187--205}.
\bid{doi={10.1111/j.1467-9868.2004.00439.x}, issn={1369-7412}, mr={2035766}}
\bptok{imsref}%
\end{barticle}
%
\endbibitem

\bibitem[\protect\citeauthoryear{Tobler}{1970}]{Tobler1970}
%
\begin{barticle}[author]
\bauthor{\bsnm{Tobler},~\bfnm{W.}\binits{W.}}
(\byear{1970}).
\btitle{A computer movie simulating urban growth in the Detroit region}.
\bjournal{Economic Geography}
\bvolume{46}
\bpages{234--240}.
\bptok{imsref}%
\end{barticle}
%
\endbibitem

\bibitem[\protect\citeauthoryear{Tucker
et~al.}{2005}]{TuckerPinzonBrownSlaybackPakMahoneyVermoteSaleous2005}
%
\begin{barticle}[author]
\bauthor{\bsnm{Tucker},~\bfnm{C.}\binits{C.}},
\bauthor{\bsnm{Pinzon},~\bfnm{J.}\binits{J.}},
\bauthor{\bsnm{Brown},~\bfnm{M.}\binits{M.}},
\bauthor{\bsnm{Slayback},~\bfnm{D.}\binits{D.}},
\bauthor{\bsnm{Pak},~\bfnm{E.}\binits{E.}},
\bauthor{\bsnm{Mahoney},~\bfnm{R.}\binits{R.}},
\bauthor{\bsnm{Vermote},~\bfnm{E.}\binits{E.}} \AND
\bauthor{\bsnm{Saleous},~\bfnm{N.}\binits{N.}}
(\byear{2005}).
\btitle{An extended AVHRR 8-km NDVI data set compatible with MODIS and SPOT
vegetation NDVI data}.
\bjournal{International Journal of Remote Sensing}
\bvolume{26}
\bpages{4485--4498}.
\bptok{imsref}%
\end{barticle}
%
\endbibitem

\bibitem[\protect\citeauthoryear{Usongo and
Nagahuedi}{2008}]{UsongoNagahuedi2008}
%
\begin{barticle}[author]
\bauthor{\bsnm{Usongo},~\bfnm{L.}\binits{L.}} \AND
\bauthor{\bsnm{Nagahuedi},~\bfnm{J.}\binits{J.}}
(\byear{2008}).
\btitle{Participatory land-use planning for priority landscapes of the
Congo Basin}.
\bjournal{Unasylva}
\bvolume{230}
\bpages{17--24}.
\bptok{imsref}%
\end{barticle}
%
\endbibitem

\bibitem[\protect\citeauthoryear{Vrieling, de~Beurs and
Brown}{2008}]{VrielingdeBeursBrown2008}
%
\begin{bmisc}[author]
\bauthor{\bsnm{Vrieling},~\bfnm{A.}\binits{A.}},
\bauthor{\bparticle{de} \bsnm{Beurs},~\bfnm{K.~M.}\binits{K.~M.}} \AND
\bauthor{\bsnm{Brown},~\bfnm{M.~E.}\binits{M.~E.}}
(\byear{2008}).
\bhowpublished{Recent trends in agricultural production of Africa based
on AVHRR NDVI
time series. \textit{Proceedings of the SPIE Conference: Remote
Sensing for Agriculture,
Ecosystems and Hydrology X}}.
\bptok{imsref}%
\end{bmisc}
%
\endbibitem

\end{thebibliography}
\end{document}